\documentclass[sigconf]{acmart}
\AtBeginDocument{%
  }

\copyrightyear{2026}
\acmYear{2026}
\setcopyright{cc}
\setcctype{by-nc-nd}
\acmConference[DAC '26]{63rd ACM/IEEE Design Automation Conference}{July 26--29, 2026}{Long Beach, CA, USA}
\acmBooktitle{63rd ACM/IEEE Design Automation Conference (DAC '26), July 26--29, 2026, Long Beach, CA, USA}
\acmDOI{10.1145/3770743.3804323}
\acmISBN{979-8-4007-2254-7/2026/07}

\usepackage{graphicx}
\usepackage{listings}
\usepackage{xcolor}
\usepackage{subcaption}
\usepackage{tabularx}
\usepackage{booktabs}
\usepackage{enumitem}

\begin{document}

\newcommand{\highlight}[1]{\textcolor{orange}{#1}}  
\newcommand{\todo}[1]{\textcolor{red}{#1}} 
\newcommand{\revised}[1]{\textcolor{blue}{#1}}
\newcommand{\nxt}[1]{\textcolor{cyan}{#1}}

\title{vTen: Tensor-Centric Verification Framework for Domain-Specific Accelerators}

\author{{Chanmin Baek, Keehyuk Lee, Mincheol Cha, Somi Hong, Xuan Truong Nguyen\footnotemark[1], Hyuk-Jae Lee}}
\email{{{bcm0314, lkh099, mccha}@capp.snu.ac.kr, ooihong0906@snu.ac.kr, {truongnx, hyuk_jae_lee}@capp.snu.ac.kr}}
\affiliation{%
  \institution{Seoul National University}
  \city{Seoul}
  \country{South Korea}
}

\renewcommand{\shortauthors}{Baek et al.}

\begin{abstract}

The semantic gap between tensor-centric software models and signal-level hardware testbenches creates significant productivity bottlenecks in verifying Domain-Specific Accelerators (DSAs). 
Existing frameworks like Cocotb suffer from prohibitive synchronization overheads due to fine-grained interactions. 
To address this, we propose vTen, a data-centric framework that strictly decouples verification intent from execution mechanics. 
By leveraging a declarative DSL and kernel-granular batching, vTen minimizes host-simulator interaction frequency. 
Evaluation on a production-scale 3D U-Net accelerator demonstrates that vTen achieves a 2$\times$ performance improvement in simulation latency and a 60.3\% reduction in code complexity compared to Cocotb.

\end{abstract}





\maketitle
\footnotetext[1]{Corresponding author: Xuan Truong Nguyen (truongnx@capp.snu.ac.kr).}

\section{Introduction} \label{sec:introduction}

Deep learning models have been widely adopted for various tasks such as image classification~\cite{he_resnet}, object detection~\cite{yolov3}, and image segmentation~\cite{unet}.
The rapid popularity of diverse AI models is driving the development of domain-specific accelerators across academia~\cite{nguyen_tcasi_2022, kim_tcasi_2024, lee_icce_2024, dfx_hcs2022, spatten} and industry~\cite{tpu, gemmini}. 
This workload diversity has driven the proliferation of domain-specific accelerators (DSAs) across the entire design stack—from RTL simulation and SoC integration to FPGA prototyping and ASIC tape-out. 
As a result, verification --- the process of validating the correctness of hardware modules (e.g., register-transfer level (RTL) designs or FPGA implementation) against golden software workloads (e.g., written in PyTorch~\cite{pytorch} or TensorFlow~\cite{tensorflow}) --- appears as an emerging challenge in hardware design development. Industry surveys report that verification consumes 60--70\% of design effort~\cite{wilson2024}.

\begin{figure}[t]
    \vspace{8pt}
    \centering
    \begin{subfigure}[b]{0.32\linewidth}
        \centering
        \includegraphics[width=\linewidth]{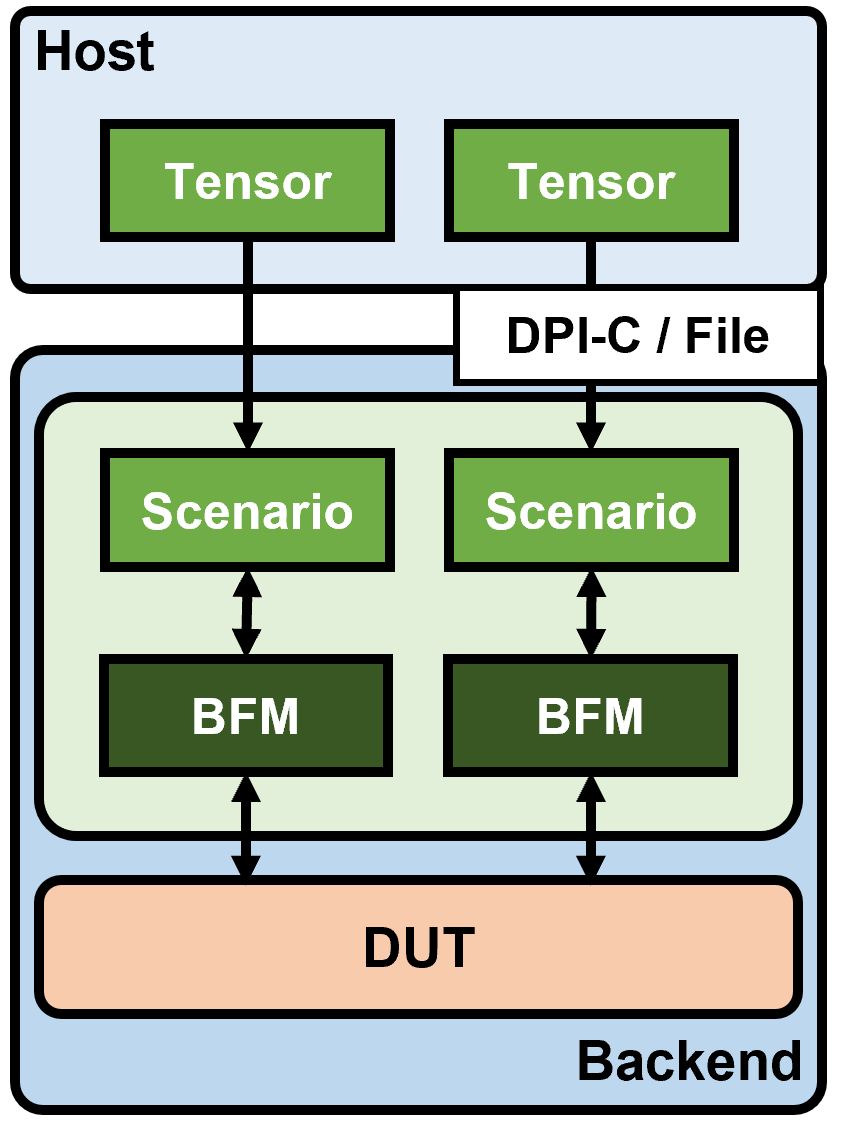}
        \caption{UVM}
        \label{fig:uvm}
    \end{subfigure}
    \hfill
    \begin{subfigure}[b]{0.32\linewidth}
        \centering
        \includegraphics[width=\linewidth]{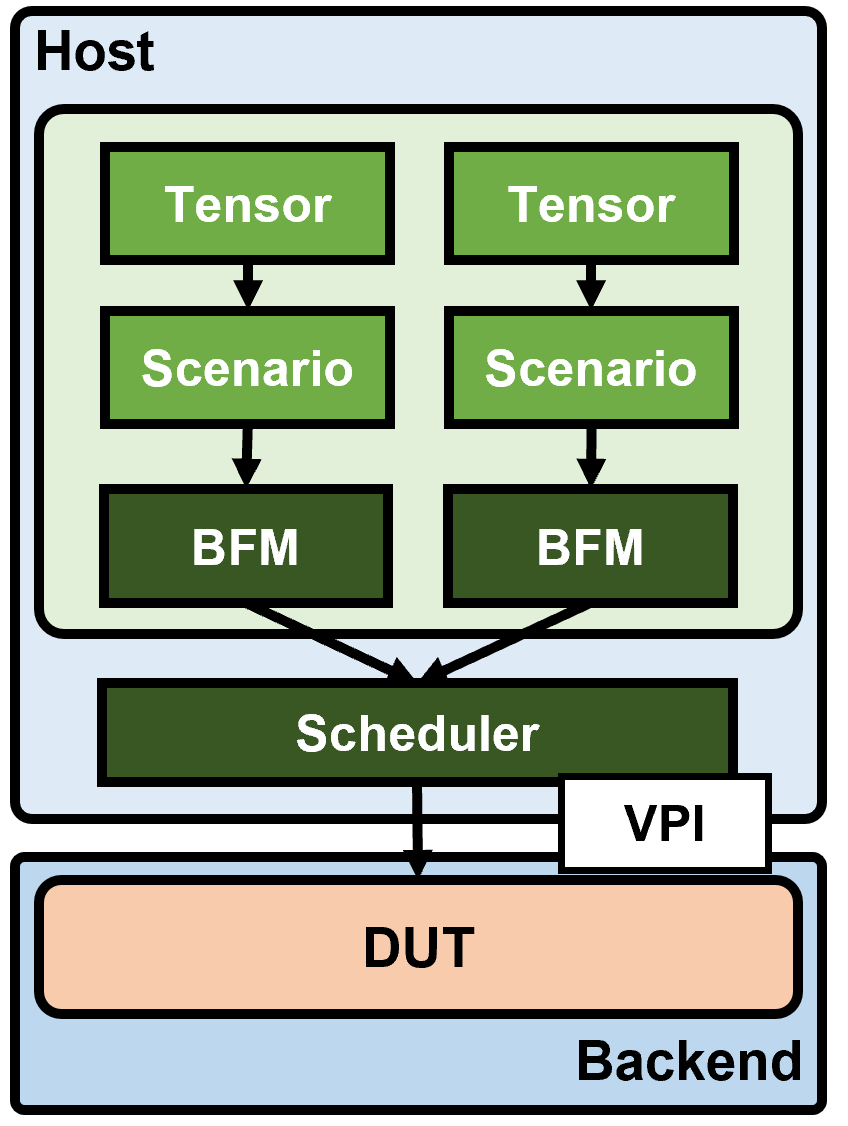}
        \caption{Cocotb}
        \label{fig:cocotb}
    \end{subfigure}
    \hfill
    \begin{subfigure}[b]{0.32\linewidth}
        \centering
        \includegraphics[width=\linewidth]{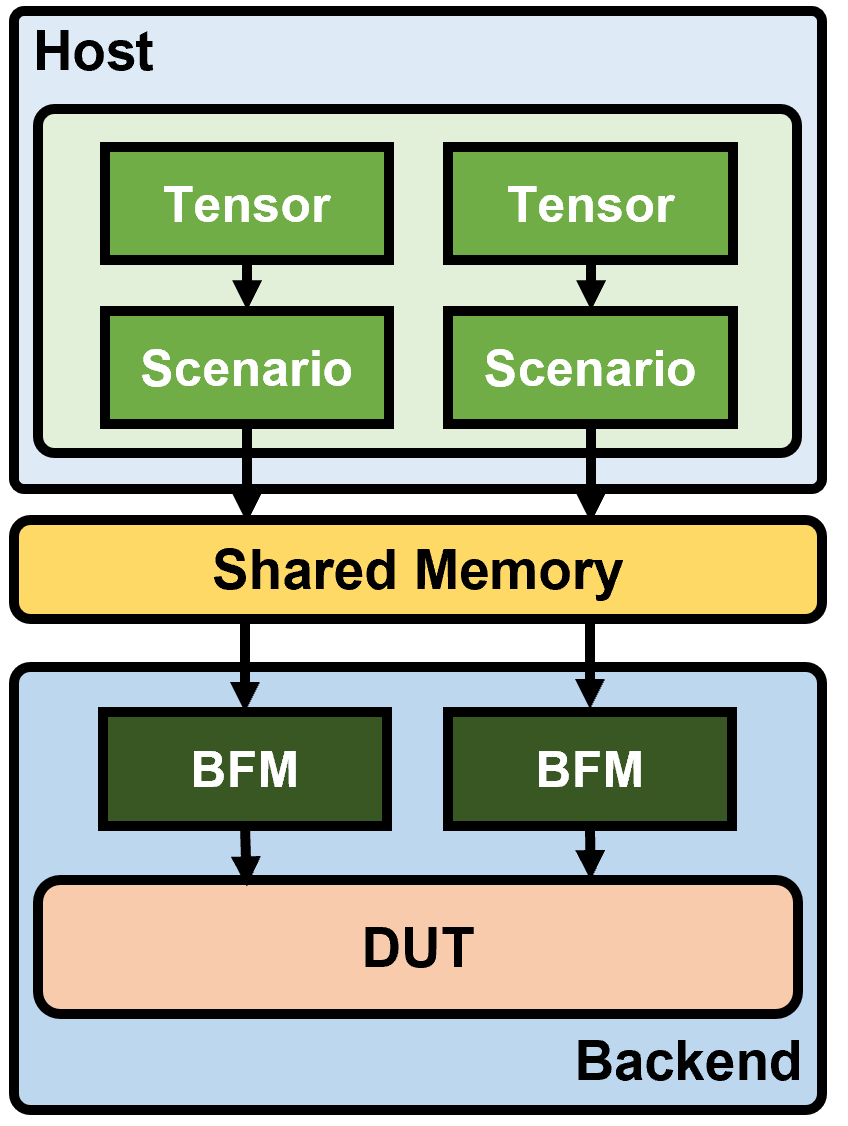}
        \caption{vTen}
        \label{fig:vten}
    \end{subfigure}
    
    \caption{Comparison of Verification Frameworks: (a) UVM, (b) Cocotb, and (c) vTen.}
    \label{fig:architecture}
    \vspace{-2pt}
\end{figure}

The growing complexity of integrated circuits has intensified the challenges in hardware design verification. While SystemVerilog and Unified Verification Methodology (UVM)~\cite{uvm} are industry standards, their complexity raises entry barriers. For example, as illustrated in Figure~\ref{fig:architecture}(a), UVM requires converting a test vector, such as a Tensor, into a file that can be transferred to a simulation backend via Direct Programming Interface (DPI-C). However, UVM testing scenarios are handled at the backend, limiting its ability to handle diverse tensor data on the host or frontend. As verification is widely accepted as a software task, leveraging high-level languages such as C++ and Python has become a trend. Current hardware verification languages such as Python in Cocotb~\cite{cocotb}, and LuaJIT in Verilua~\cite{verilua} offer promising solutions to handle diverse cases in the frontend. For instance, as shown in Figure~\ref{fig:architecture}(b), testing scenarios are coupled with data tensors and a bus functional model (BFM) on the frontend, and the controls are transferred to the backend via a Verilog Procedural Interface (VPI). However, they may require transferring a large number of transactions via VPI, which typically results in slow simulation performance. Additionally, the approaches generally face poor component reusability across simulation environments.

\setlength{\textfloatsep}{6pt plus 1pt minus 1pt}
\setlength{\floatsep}{6pt plus 1pt minus 1pt}
\setlength{\intextsep}{6pt plus 1pt minus 1pt}
\setlength{\abovecaptionskip}{9pt}
\setlength{\belowcaptionskip}{-8pt}

A typical hardware verification flow involves translation across multiple levels, including tensors, bytes, transactions, and control signals, combining both data-centric and control-centric layers.
As illustrated in Figure~\ref{fig:architecture}(c), a data-centric layer includes tensors and bytes that are primarily associated with data, its format, and testing scenarios for functional correctness verification on the software side.
In contrast, a control-centric layer relates to the simulation backend, which primarily handles sending bus functional transactions and control signals on the hardware side.
This observation poses a fundamental challenge in hardware design verification for DSAs, which is \emph{a semantic gap}. 
Specifically, the essential goal of the data-centric layer differs from that of the control-centric layer. 
While the data-centric approach aims to verify the functional correctness of a hardware module by covering diverse test data, the control-centric layer primarily integrates simulation and waveform analysis.
This suggests a strong message that an inappropriate coupling of them can be considered a root cause of productivity bottlenecks and performance overheads in existing methodologies like~\cite{uvm, cocotb}.

To overcome these challenges, we propose vTen, a data-centric verification framework that bridges a PyTorch-based tensor-centric dataflow and the hardware backend simulation. 
The main contributions of this work are as follows.
\vspace{-1pt}
\begin{itemize}
    \item \textbf{Methodology.} We propose a Data-Centric Abstraction that strictly decouples verification intent from execution mechanics. By combining a tensor-based Domain Specific Language (DSL) with Kernel-Granular Execution, vTen automates the translation of high-level dataflow into batched, protocol-specific transactions to minimize synchronization overhead.

    \item \textbf{Architecture.} To realize this methodology, vTen introduces a layered architecture that consists of a \textit{Frontend} for defining verification scenarios using a DSL, a \textit{Runtime} that lowers these scenarios into a backend-agnostic Execution Intermediate Representation, and a split-process \textit{Backend} that executes batched operations via shared memory. This structure effectively bridges the semantic gap while maximizing simulation throughput.
    
    \item \textbf{Evaluation.} We validate vTen on a 3D CNN accelerator across diverse workloads, demonstrating: (1) 2$\times$ speedup in test execution compared to Cocotb through kernel-granular batching, (2) a 60.3\% reduction in verification code complexity via the declarative DSL.
\end{itemize}

\section{Background and Motivation}\label{sec:background}

\subsection{AI Domain-Specific Accelerators} \label{subsec_DSA}
Modern deep learning models typically consist of general matrix multiplication (GEMM) operations (e.g., convolution or fully connected layers) and non-GEMM operations, such as memory operations, normalization, and logit computation. 
Among them, GEMM operations are still dominant in terms of the number of operations and overall latency, ranging from 42.8\% to 96.6\%~\cite{karami_ispass25_nongemm}, and have been the prime optimization targets for domain-specific accelerators. 
The fundamental computing pattern of GEMM operations is the large use of basic operations such as multiplication and accumulation (MAC), which can be accelerated by massive parallel processing elements (PEs)~\cite{nguyen_tcasi_2022, kim_tcasi_2024, lee_icce_2024, dfx_hcs2022, spatten, tpu, gemmini}.

To leverage highly parallel PEs, it is essential to consider massive data movements across various memory levels, including HBM or DRAM, cache or global memory, and registers~\cite{nguyen_tcasi_2022, kim_tcasi_2024, tpu}. 
For example, to execute a convolution layer, a DSA transfers weights and input feature maps (IFMs) from an external memory to a global on-chip buffer. 
In a DSA, data is moved between buffers and a PE array via a standard bus interface, such as AXI. 
The output results (so-called output feature maps (OFMs)) can be stored in on-chip memory or transferred to external memory via a direct memory access (DMA) module. 
In short, designing an AI accelerator requires managing massive data movements across different modules.

\subsection{UVM and Cocotb}
UVM~\cite{uvm} and Cocotb~\cite{cocotb} present two essential approaches for hardware design verification. UVM is generally convenient for hardware designers by enabling low-level access and manipulation of test scenarios and parameters in the low-level hardware description language (HDL). Meanwhile, Python-based Cocotb is particularly appealing to computer architects who typically use cycle-level (CL) modeling methodologies~\cite{pymtl3}. Such CL-based approaches provide significant productivity benefits. For example, in a development flow with continuous integration, they allow for gradually replacing existing CL blocks with newly developed RTL designs in a large design, thanks to its abstraction.

\subsection{Challenges for DSA Verification}
DSA verification presents a unique challenge for existing methodologies such as UVM~\cite{uvm} and Cocotb~\cite{cocotb}. As explained in Section~\ref{sec:introduction}, there exists a fundamental semantic gap between data-centric testbenches and control-centric hardware simulation during verification. The gap can be intensified when verifying hardware modules in a DSA design. 
The challenges can be described as follows.

\setlength{\textfloatsep}{6pt plus 1pt minus 1pt}
\setlength{\floatsep}{6pt plus 1pt minus 1pt}
\setlength{\intextsep}{6pt plus 1pt minus 1pt}
\setlength{\abovecaptionskip}{4pt}
\setlength{\belowcaptionskip}{-9pt}
\begin{figure}[t]
    \centering
    \lstset{
        basicstyle=\ttfamily\scriptsize,
        frame=single,
        numbers=none,
        xleftmargin=0.5em,
        breaklines=true,
        escapechar=|,
        keywordstyle=\color{blue},
        commentstyle=\color{gray},
        aboveskip=0em,
        belowskip=0.2em
    }

    \begin{minipage}[t]{0.48\linewidth}
        \textbf{(a) UVM (Hybrid flow)} \\[-1em]
        
        \begin{lstlisting}[language=Python, title={\textbf{Data Gen: Python}}]
# Same setup
fm = torch.randn(D, C, H, W)
packed_data = Pack(Quant(fm))
# Send to UVM via DPI/Socket
uvm_conn.send(packed_data)
        \end{lstlisting}
        
        \vspace{-0.1cm}
        \begin{center}
            $\Downarrow$ \scriptsize \textit{Cross-Language Transfer}
        \end{center}
        \vspace{-0.1cm}

        \begin{lstlisting}[language=Verilog, title={\textbf{Driver: SystemVerilog}}]
// Native Execution Loop
task run_phase(uvm_phase phase);
  forever begin
    seq_item_port.get(req);
    
    // Pin Wiggling (Fast)
    vif.tdata  <= req.data;
    vif.tvalid <= 1'b1;
    
    @(posedge vif.clk);
    while(!vif.tready) @(posedge vif.clk);
    seq_item_port.item_done();
  end
endtask
        \end{lstlisting}
    \end{minipage}
    \hfill
    \begin{minipage}[t]{0.48\linewidth}
        \textbf{(b) Cocotb (Python-driven)} \\[-1em]
        
        \begin{lstlisting}[language=Python, title={\textbf{Data Gen: Python}}]
# Same setup
fm = torch.randn(D, C, H, W)
packed_data = Pack(Quant(fm))

|\phantom{}|
        \end{lstlisting}

        \vspace{-0.1cm}
        \begin{center}
            $\Downarrow$ \scriptsize \textit{Internal Function Call}
        \end{center}
        \vspace{-0.1cm}

        \begin{lstlisting}[language=Python, title={\textbf{Driver: Python}}]
# High Overhead Loop
for word in packed_data:
    # VPI Write
    dut.tdata.value = word
    # VPI Write
    dut.tvalid.value = 1

    # VPI Wait / Context Switch
    await RisingEdge(dut.clk)
    # VPI Read
    while dut.tready.value == 0:
        # VPI Write
        await RisingEdge(dut.clk)
|\phantom{}|
        \end{lstlisting}
    \end{minipage}
    \caption{An example of simulating a tensor with (a) UVM and (b) Cocotb.}
    \label{fig:cocotb_vs_uvm}
\end{figure}
\begin{figure*}[t]
    \centering
    \includegraphics[width=\textwidth]{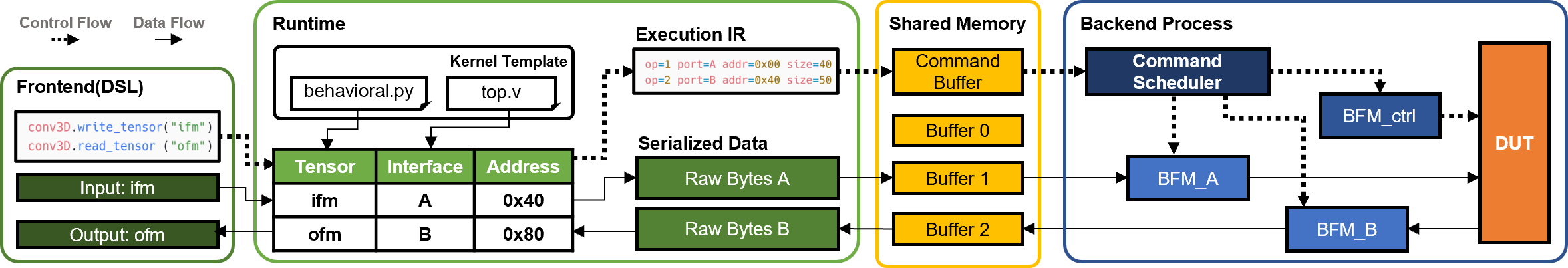}
    \caption{Overall Architecture of vTen.}
    \label{fig:overall_arch}
\end{figure*}
\noindent
\textbf{Challenge 1 (Massive Repetitive Operations)}: To validate the functional correctness of a module in a DSA, it is generally necessary to conduct a simulation with large-scale tensors, for example, weights or IFMs of multiple convolution layers. As a result, it may impose considerable burdens on hardware designers when adopting UVM~\cite{uvm} for simulation/verification of large-scale tensors. 
For example, as illustrated in Figure~\ref{fig:cocotb_vs_uvm}(a), with UVM, the driving loop is offloaded to SystemVerilog. However, it requires complex cross-language data transfer, such as \texttt{uvm\_conn.send(packed\_data)}.

Meanwhile, from a control-centric simulation perspective, the flow is highly repetitive, consisting of a set of predefined operations. 
Here, as demonstrated in Figure~\ref{fig:cocotb_vs_uvm}(b), with Cocotb, the iteration occurs in the Python runtime (\texttt{for word in packed\_data}).
Consequently, every single data beat triggers multiple VPI operations: writing values (\texttt{dut.tdata.value}), scheduling callbacks (\texttt{await RisingEdge}), and reading status (\texttt{dut.tready}). 
For DNN workloads involving millions of cycles, this chatter between Python and the backend process creates a prohibitive performance bottleneck.

\noindent
\textbf{Challenge 2 (System Integration)}: Designing a DSA also involves integrating and validating multiple block designs, including massive-parallel PEs, global buffers, line buffers, DMAs, and controllers. 
In this context, UVM~\cite{uvm} suffers from a critical limitation in its ability to abstract block designs, especially for the verification of functional correctness. 
In contrast, using Cocotb~\cite{cocotb} generally prolongs simulation time due to substantial VPI overhead at every sub-module in a large-scale design.

These observations pose a strong demand to design an alternative verification framework.
A target design must effectively bridge the semantic gap between tensors and hardware simulation.
It is also essential to achieve scalable performance for data-intensive workloads while achieving portability for design reuse and system integration.
\section{vTen Architecture} \label{sec_vTen}

\subsection{Methodology and Architecture Overview}
\textbf{Methodology.} vTen bridges the semantic gap and mitigates synchronization overheads via a tensor-centric approach, prioritizing throughput without sacrificing flexibility for diverse DSAs.
The design rests on three fundamental principles: 
(1) \emph{Data-Centric Abstraction} --- decouples verification intent from execution mechanics by elevating the specification level from pin-level transactions to operations on tensors and registers;
(2) \emph{Kernel-Granular Execution} --- treats the \textit{Kernel}---defined as the accelerator's top-level DUT encompassing compute logic and interfaces--- as the atomic unit of interaction. By batching tensor operations into a single execution unit, it reduces host-simulator synchronization frequency from $O(C)$ to $O(1)$, where $C$ denotes the kernel execution cycles;
(3) \emph{Shared Memory Data Plane} --- utilizes serialized byte streams to enable direct backend access, bypassing the high-latency IPC overhead inherent in transaction-level interfaces.

\noindent 
\textbf{Architecture.} Following the design methodology, vTen introduces a layered architecture, as illustrated in Figure~\ref{fig:overall_arch}.
Instead of a monolithic testbench, the framework decouples verification into three distinct layers, effectively separating high-level intent from low-level execution mechanics. 
First, at the \textbf{Frontend or Intent layer}, users define verification intent using a Python-based DSL. This layer focuses purely on dataflow, abstracting away cycle-level timing and physical port protocol. 
Second, at the \textbf{Runtime or Abstraction layer}, the runtime engine leverages Kernel Abstraction definitions to automatically lower high-level DSL operations into a backend-agnostic Execution Intermediate Representation (Execution IR) while handling tensor serialization and address mapping. 
Third, at the \textbf{Backend or Execution layer}, a split-process execution engine executes the compiled Execution IR batches. By utilizing shared memory and batch execution, the backend maximizes simulation throughput while reducing host-simulator transactions.

The subsequent subsections detail the core components corresponding to these layers: 
Section~\ref{sec:dsl} introduces the DSL for intent specification;
Section~\ref{sec:kernel_abstraction} describes the Kernel Abstraction mechanisms managed by the runtime;
and Section~\ref{sec:execution_layer} explains the split-process Execution Layer that powers the backend.

\subsection{Domain-Specific Language for Data-Centric Verification} \label{sec:dsl}

\subsubsection{Core Abstractions.}
The DSL serves as the declarative frontend interface, enabling users to define verification intent through high-level dataflow rather than low-level signaling.
It provides a set of hardware-agnostic primitives that act as abstract templates for data movement (e.g., \texttt{write\_tensor}, \texttt{read\_register}) and logical synchronization (e.g., \texttt{barrier}). 
By abstracting away specific transport mechanisms (e.g., AXI4-Stream vs. AXI4-MM), these primitives strictly decouple the verification logic from interface implementation details. 
This allows the testbench to focus solely on the tensor-level dataflow intent, leaving the Runtime to automatically bind these abstract operations to the specific hardware protocols via Kernel Abstraction (Section 3.3).
\vspace{-6pt}

\setlength{\textfloatsep}{4pt plus 1pt minus 1pt}
\setlength{\floatsep}{6pt plus 1pt minus 1pt}
\setlength{\intextsep}{6pt plus 1pt minus 1pt}
\setlength{\abovecaptionskip}{4pt}
\setlength{\belowcaptionskip}{-6pt}
\begin{table}[t]
    \vspace{4pt}
    \centering
    \renewcommand{\arraystretch}{1}
    \caption{Portable DSL Operations (Level 1)}
    \label{tab:dsl_ops}
    \small
    \begin{tabularx}{\columnwidth}{@{} l @{\hspace{2em}} X @{}}
        \toprule
        \textbf{Operation} & \textbf{Description} \\
        \midrule
        \texttt{write\_tensor} & Transfers tensor data to DUT. \\
        \texttt{read\_tensor} & Retrieves tensor data from DUT. \\
        \texttt{write\_register} & Configures control register. \\
        \texttt{read\_register} & Queries current register state. \\
        \texttt{poll\_register} & Blocks until register satisfies predicate. \\
        \texttt{barrier} & Enforces global synchronization. \\
        \bottomrule
    \end{tabularx}
    \vspace{-2pt}
\end{table}

\subsubsection{Parallelism and Dependency Model.} Hardware designs inherently exhibit massive spatial parallelism and non-blocking interactions. 
Expressing these concurrent behaviors within a sequential DSL syntax—without resorting to complex explicit constructs like SystemVerilog's \texttt{fork-join}—presents a semantic challenge. 
To reconstruct this concurrency from a sequential DSL, vTen defines the operation lifecycle using two distinct events, \textit{Issue} and \textit{Commit}. 
Specifically, \textit{Issue} marks the time stamp when an operation is dispatched to the DUT, while \textit{Commit} marks the moment the DUT actually finishes execution and returns the result. 
The timing of these events is managed through the following two mechanisms:

\begin{enumerate}[label=(\arabic*), leftmargin=*, noitemsep]
    \item \textbf{Synchronization Mode (Implicit Control):} 
    \textit{Async} ($Issue_{n} \approx Issue_{n+1}$) allows immediate dispatch for parallel pipelines, where\-as \textit{Sync} enforces serial execution ($Commit_{n} \prec Issue_{n+1}$) by stalling the issue of the subsequent operation.

    \item \textbf{Explicit Dependencies (Fine-grained Control):} 
    Basic concurrency is refined by \textit{Issue Dependencies} ($Commit_{src} \prec Issue_{dst}$) to resolve data hazards (e.g., RAW), and \textit{Commit Dependencies} ($Commit_{dst} \preceq Commit_{src}$) to model dynamic termination events like interrupts.

    \item \textbf{Barrier Primitive (Global Synchronization):}
    To enforce coarse-grained ordering, the \texttt{barrier} primitive acts as a global fence. It stalls the dispatch of any subsequent operations until all in-flight operations reach their \textit{Commit} state ($Commit_{\forall prev} \prec Issue_{next}$).
\end{enumerate}
\vspace{-8pt}

\setlength{\textfloatsep}{6pt plus 1pt minus 1pt}
\setlength{\floatsep}{6pt plus 1pt minus 1pt}
\setlength{\intextsep}{6pt plus 1pt minus 1pt}
\setlength{\abovecaptionskip}{5pt}
\setlength{\belowcaptionskip}{-4pt}
\begin{figure}[t]
    \centering
    \begin{subfigure}[b]{0.55\linewidth}
        \begin{lstlisting}[language=Python, basicstyle=\ttfamily\scriptsize, breaklines=true, columns=fullflexible, numbers=left, xleftmargin=1.5em]
# Layer N: Synchronization point
w0 = ctx.write_register("ctrl", 0x1)

# Branch 1: Process feature map 1
w1 = ctx.write_tensor("ifm_1", dep=w0)
r1 = ctx.read_tensor("ofm_1", dep=w1)

# Branch 2: Monitor completion
p2 = ctx.poll_register("done", dep=w0)

# Commit Dependency(r1, p2)
r1.add_dependency(p2)

# Layer N+1: Next synchronization
w3 = ctx.write_register("ctrl", dep=r1)
        \end{lstlisting}
        \caption{DSL Specification}
        \label{fig:code}
    \end{subfigure}
    \hfill
    \begin{subfigure}[b]{0.4\linewidth}
        \includegraphics[width=\linewidth]{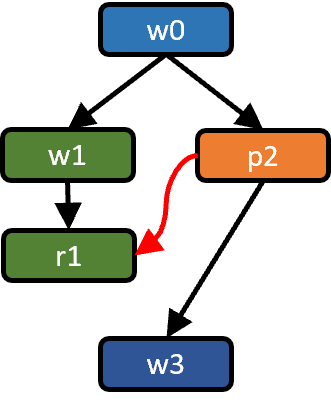}
        \caption{Dataflow Graph}
        \label{fig:image}
    \end{subfigure}
    
    \caption{Example of mapping DSL code (a) and dataflow graph (b).}
    \label{fig:dsl_example}
\end{figure}

\subsubsection{Example: Parallel Feature Map Processing.}

Figure~\ref{fig:dsl_example} demonstrates a fork-join pattern implemented via the DSL. 
The sequence initiates with a blocking operation $w_0$ (Line 2). 
Upon its completion, the control flow \textit{diverges}: the primary data path executes the sequence $w_1 \to r_1$, while a background polling task $p_2$ is issued concurrently via \textit{Async} mode.
Line 12 establishes a \textit{Commit Dependency} ($Commit_{r_1} \preceq Commit_{p_2}$), acting as a termination bound to preemptively terminate $r_1$ upon $p_2$'s completion.
The two branches \textit{converge} at the synchronization point $w_3$ (Line 15), restoring serial execution.

\subsection{Bridging the Semantic Gap: Kernel Abstraction} \label{sec:kernel_abstraction}
To bridge the semantic gap, the \textbf{Kernel Abstraction} layer provides the necessary mapping intelligence to drive physical hardware ports from high-level tensor specifications.
It automates the lowering of DSL operations into address-aware \textit{Execution IR} and manages the data serialization/deserialization processes, thereby decoupling verification logic from architectural details. 
This abstraction is materialized through a two-phase lifecycle:

\subsubsection{Build-time Kernel Template.}
The Kernel Template serves as a static, parameterizable blueprint of the Kernel.
It is constructed by extracting interface specifications from lightweight RTL pragmas and retrieving tensor/register definitions from the Python-based functional reference model.
Architectural attributes---including tensor shapes, register reset values, and protocol parameters---can remain parameterized at this stage.
This separation allows verification logic to remain agnostic to specific architectural tuning (e.g., changing array sizes or bus widths).

\setlength{\textfloatsep}{6pt plus 1pt minus 1pt}
\setlength{\floatsep}{6pt plus 1pt minus 1pt}
\setlength{\intextsep}{6pt plus 1pt minus 1pt}
\setlength{\abovecaptionskip}{5pt}
\setlength{\belowcaptionskip}{0pt}
\begin{figure}[t]
    \centering
    \lstset{
        basicstyle=\ttfamily\scriptsize, 
        frame=single,
        numbers=none,
        breaklines=true,                 
        breakatwhitespace=false,         
        postbreak=\mbox{\textcolor{gray}{$\hookrightarrow$}\space},
        breakindent=1em,                 
        xleftmargin=0.5em,
        columns=fullflexible            
    }

    \lstset{
        basicstyle=\ttfamily\scriptsize,
        frame=single,
        breaklines=true,
        escapechar=|,  
        numbers=none,
        aboveskip=0pt
    }
    
    \begin{minipage}[t]{0.48\linewidth}
        \textbf{(a) Interface Spec (RTL)} \\
        \begin{lstlisting}[language=Verilog]
// vten:interface ifm_axi axi4_master
// vten:interface:parameter [addr: 64, data: 256]
// vten:interface:addr_map ddr
// vten:interface:tensor ifm unsigned[8]
        \end{lstlisting}
    \end{minipage}
    \hfill
    \begin{minipage}[t]{0.48\linewidth}
        \textbf{(b) Functional.py} \\
        \begin{lstlisting}[language=Python]
self.ifm = Tensor(
    shape=(1, "${C}", 32, 32),
    dtype=torch.float32)
def forward(self): 
    return F.conv2d(self.ifm, self.weight)
        \end{lstlisting}
    \end{minipage}
    
    \begin{center}
        \large $\Big\downarrow$ \small \textit{Automated Binding}
    \end{center}
    \vspace{0.1cm}

    \begin{minipage}[t]{1.0\linewidth}
        \textbf{(c) Constructed Binding Table (Runtime)} \\
        \begin{lstlisting}[language=Python]
TensorBinding(
    tensor_name="ifm",             # Linked from functional model's tensor
    interface_name="ifm_axi",      # Linked from RTL interface
    resolved_shape=(1, 64, 32, 32),# ${C} resolved to 64
    serializer=DataSerializer(...),
    address_mapping=AddressMapping(base=0x0, size=131072, addr_map="ddr")
)
        \end{lstlisting}
    \end{minipage}

    \caption{\textbf{Overview of the vTen compilation flow.} Parameters and interfaces from RTL (a) and Python (b) are automatically resolved and bound in the runtime table (c).}
    
    \label{fig:binding_flow}
    \vspace{-1em}
\end{figure}

\subsubsection{Runtime Kernel Object.} 
At runtime, the Kernel Template is instantiated into a concrete Kernel Object. 
The primary objective of this phase is to construct the \textit{Binding Table}, which serves as the definitive logical-to-physical map for execution, as illustrated in Figure~\ref{fig:binding_flow}. 
This is constructed through three consolidated steps.

First, the runtime concretizes the kernel configuration by resolving parameterized expressions within a unified namespace.
By binding abstract attributes (e.g., \texttt{\$\{C\}}) to target-specific values (e.g., \texttt{\{C: 64\}}), vTen enables seamless design space exploration across different accelerator variants without code modification.
Second, the runtime establishes a bidirectional pipeline to bridge the software-hardware format gap.
For serialization, the pipeline executes a three-stage transformation: (i) \textit{Layout Transformation} to match the hardware memory hierarchy; (ii) \textit{Quantization} to hardware fixed-point formats; and (iii) \textit{Bit-packing} into bus-aligned byte streams.
This infrastructure also supports the inverse deserialization flow, enabling the automated reconstruction of high-level tensors from raw hardware outputs.
Finally, the runtime performs physical address allocation based on the size of the serialized byte streams and the interface address map.
These resolved physical constraints are populated into the \textit{Binding Table}, enabling the execution engine to resolve any tensor access with $O(1)$ lookup overhead.
\vspace{-15pt}

\subsubsection{Execution IR Lowering and Portability.} Using the constructed \textit{Binding Table}, the runtime translates DSL primitives into Execution IR---a backend-agnostic instruction format. 
For instance, a \texttt{write\_tensor} DSL primitive is compiled into a generic command: \texttt{WRITE(interface\_id, buffer\_id, protocol, phys\_addr, dep)}. 
This intermediate representation completely decouples verification intent from the backend execution layer. 
While current backends map Execution IR to VPI/DPI calls for simulation, the architecture is explicitly designed to support hardware emulation platforms (e.g., FPGA). 
The shared memory transport model (detailed in Section 3.4) maps naturally to PCIe DMA, allowing future backends to execute the same instructions on FPGAs, ensuring test portability across the verification lifecycle.

\subsection{Scalable Execution Layer} \label{sec:execution_layer}
We leverage the Backend Execution Engine (Figure~\ref{fig:overall_arch}) to implement a scalable execution strategy. 
This architecture is designed around two core principles: minimizing runtime overhead through build-time code generation and eliminating synchronization bottlenecks via decoupled data/control transport planes.

\subsubsection{Lightweight Runtime Architecture.}
To ensure the Backend Interface operates with minimal latency, vTen enforces a strict separation between static definition and dynamic execution.
Computationally expensive tasks such as interface parsing and BFM code generation are offloaded to the \textit{Build-Time} phase.
Consequently, the \textit{Runtime} execution loop becomes extremely lightweight, handling only $O(1)$ \textit{Binding Table} lookups and dispatching pre-compiled Execution IR. 
This design eliminates recompilation overhead, facilitating rapid iterative testing on the target hardware.

\subsubsection{Split-Process Execution Model.}
To overcome the synchronization bottlenecks identified in Section 2, vTen redefines the abstraction boundary between the host and the backend. Instead of tight cycle-level coupling, we decouple execution into two distinct planes: a Data plane and a Control plane.

\noindent \textbf{Data Plane: Shared Memory Transport.} Serialized tensors are pre-populated into a shared memory data plane prior to kernel execution. 
This design isolates data transport costs from simulation logic and enables direct backend access without IPC serialization overhead.
Crucially, this transport is backend-agnostic; the backend can retrieve data either via direct pointer access (for C++ environments) or via DPI calls (for SystemVerilog simulators).
This direct access capability allows vTen to approach the theoretical performance limit of native C++ testbenches.

\noindent \textbf{Control Plane: Kernel-Granular Batch Execution.} The control plane amortizes IPC latency through Kernel-Granular Batching. 
A Kernel Task encapsulates a complete execution unit: the target kernel instance, and the compiled Execution IR sequence.
The host submits this atomic batch to the backend and the synchronization is managed via a simple lightweight handshake (Host \texttt{start} $\leftrightarrow$ Backend \texttt{complete}).
This mechanism reduces the interaction overhead from linear with cycles ($O(C)$) to constant per kernel launch $O(1)$, preserving cycle-level debugging capability via BFMs.

\subsubsection{Performance Modeling.}
To quantify the impact of the proposed architecture, we decompose the total verification latency ($T_{total}$) into four distinct stages:
(1) \textbf{Tensor Serialization ($T_{ser}$):} The runtime overhead for configuration resolution and data serialization;
(2) \textbf{Data Loading ($T_{load}$):} The transport latency for populating the shared memory buffer;
(3) \textbf{Test Execution ($T_{exec}$):} The backend simulation time dominated by the DUT's kernel execution; and
(4) \textbf{Verification ($T_{ver}$):} The post-execution phase involving output deserialization and golden result comparison.
The total latency is modeled as:
    \begin{equation}
        T_{total} = \sum_{i=1}^{N} (T_{ser} + T_{load} + T_{exec} + T_{ver})
    \end{equation}
    \noindent where $N$ denotes the total number of kernel launches in the workload.
In traditional cycle-driven frameworks (e.g., Cocotb), data transport is interleaved with execution steps, causing the combined overhead of signal driving and synchronization to scale linearly with execution cycles ($O(C)$).
In contrast, vTen's split-process architecture confines $T_{ser}$ and $T_{load}$ to the host-side runtime and utilizes batching to render synchronization overhead negligible, allowing $T_{exec}$ to dominate the total runtime ($T_{total} \approx T_{exec}$).
\section{Evaluation}

\subsection{Experimental Setup}

\setlength{\textfloatsep}{6pt plus 1pt minus 1pt}
\setlength{\floatsep}{6pt plus 1pt minus 1pt}
\setlength{\intextsep}{6pt plus 1pt minus 1pt}
\setlength{\abovecaptionskip}{3pt}
\setlength{\belowcaptionskip}{-4pt}
\begin{figure}[t]
    \centering
    \includegraphics[width=0.95\linewidth]{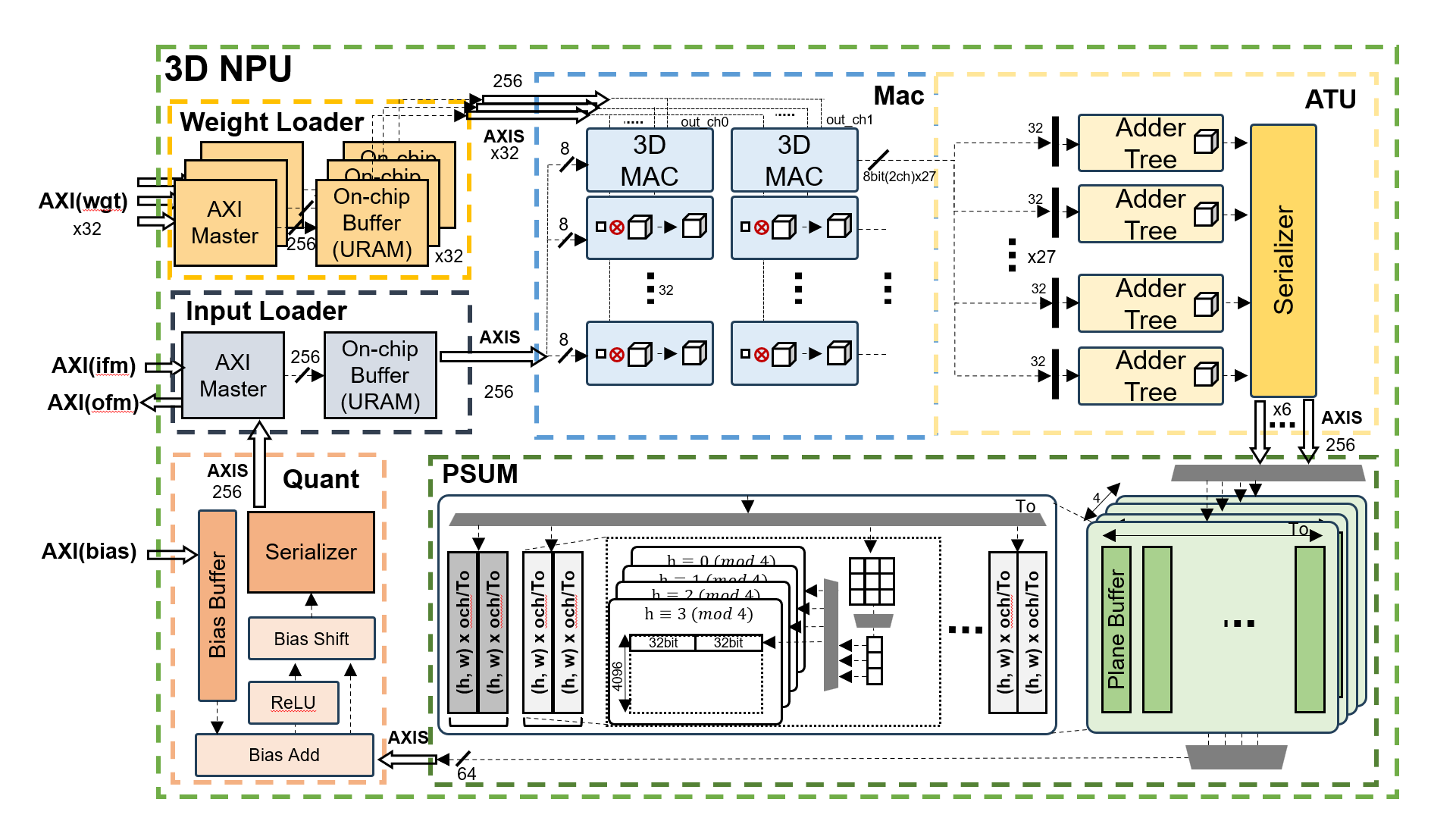} 
    \caption{An Overview of the DUT architecture. }
    \label{fig:dut_architecture}
    \descriptionlabel{}
\end{figure}

\textbf{Design Under Test (DUT).}
We evaluate the performance of vTen against Cocotb on a custom NPU shown in Figure~\ref{fig:dut_architecture} for 3D segmentation models such as 3D U-Net~\cite{20163dunet}. Such a complicated DUT is chosen to analyze Cocotb's overall VPI overhead associated with system integration of multiple submodule layers.
The architecture comprises distributed IPs linked via AXI4-Stream, utilizing AXI4-Lite for configuration and AXI4 for tensor access (IFMs, Weights, OFMs). 
These interfaces connect to vTen or Cocotb BFMs during simulation.
The design demonstrates production-scale complexity, utilizing 1,802 DSPs and 385 URAM blocks when synthesized on a Xilinx Alveo U280 FPGA at 125 MHz.

\noindent \textbf{Workloads.}
We select Conv3D layers (kernel=3, stride=1) for workload evaluation, as their large tensor footprints heavily utilize the DUT’s interface bandwidth. 
We evaluate multiple test cases with increasing tensor sizes and NPU cycle counts, starting from $(C,D,H,W)=(32,4,4,4)$ and doubling dimensions, to monitor overheads with diverse VPI count and payload sizes.

\noindent \textbf{Per-Stage Latency Analysis.}
Following Section 3.4, we decompose simulation latency into: Tensor Serialization ($T_{ser}$), Data Loading ($T_{load}$), Test Execution ($T_{exec}$), and Verification ($T_{ver}$). 
We focus on $T_{load}$ and $T_{exec}$, which directly expose architectural trade-offs, while $T_{ser}$ and $T_{ver}$ are framework-independent.

\noindent \textbf{Environment.}
Experiments run on Ubuntu 20.04 with Intel Xeon Silver 4210R (40 cores, 2.40 GHz, 1 TB RAM) using Verilator as the simulation backend.

\subsection{Abstraction Boundary Analysis}
We evaluate five transport mechanisms in increasing overhead order:
\textbf{(1) Native C++ (Theoretical Lower Bound):} A cycle-accurate C++ testbench with internal data generation, serving as the theoretical latency lower bound. 
\textbf{(2) vTen (Pure C++):} High-performance mode in which the backend reads Shared Memory (SHM) directly, bypassing SystemVerilog. 
\textbf{(3) vTen (SV-DPI):} Compatibility mode for standard simulators where SV BFMs retrieve SHM data via DPI calls. 
\textbf{(4) File I/O:} Legacy SV testbench reading vectors from disk. 
\textbf{(5) VPI (Cocotb):} Excluded from overhead analysis due to prohibitive latency ($>100\times$).

\setlength{\textfloatsep}{6pt plus 1pt minus 1pt}
\setlength{\floatsep}{6pt plus 1pt minus 1pt}
\setlength{\intextsep}{6pt plus 1pt minus 1pt}
\setlength{\abovecaptionskip}{2pt}
\setlength{\belowcaptionskip}{-8pt}
\begin{figure}[t]
    \begin{subfigure}[t]{0.48\linewidth}
        \centering
        \includegraphics[width=\textwidth]{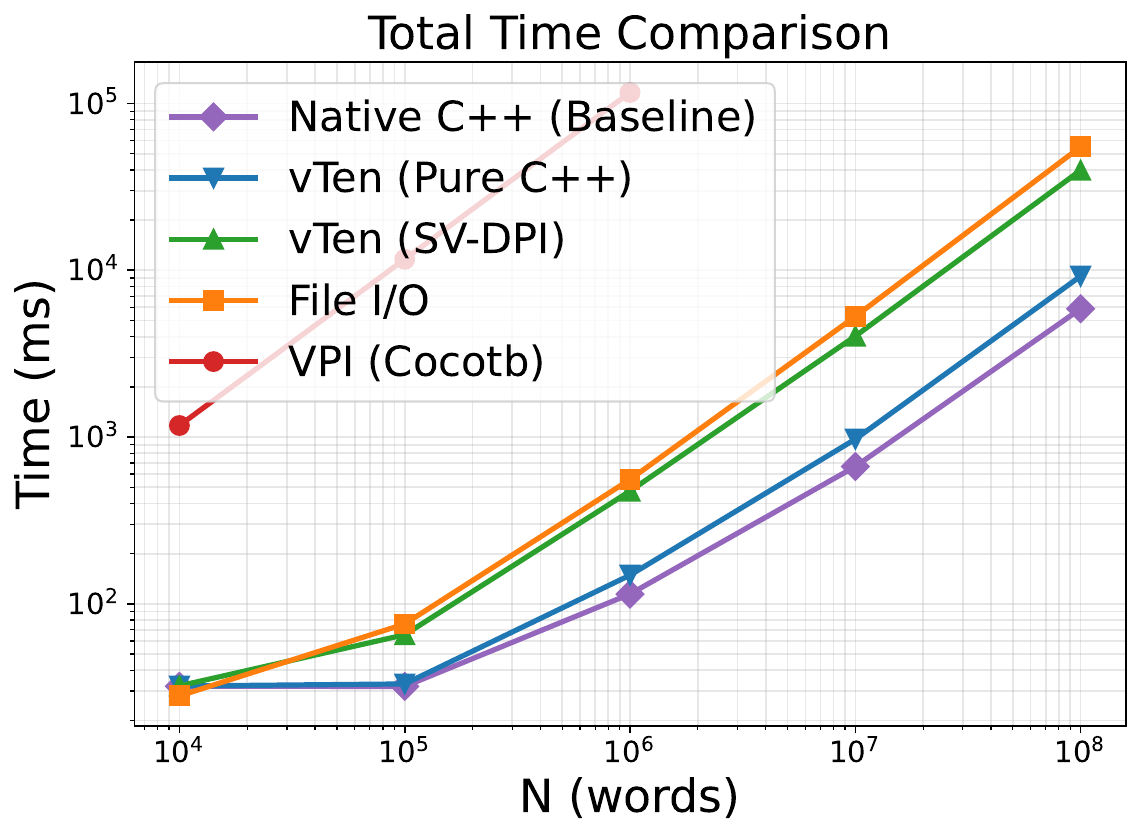} 
        \label{fig:total}
    \end{subfigure}
    \hfill
    \begin{subfigure}[t]{0.48\linewidth}
        \centering
        \includegraphics[width=\textwidth]{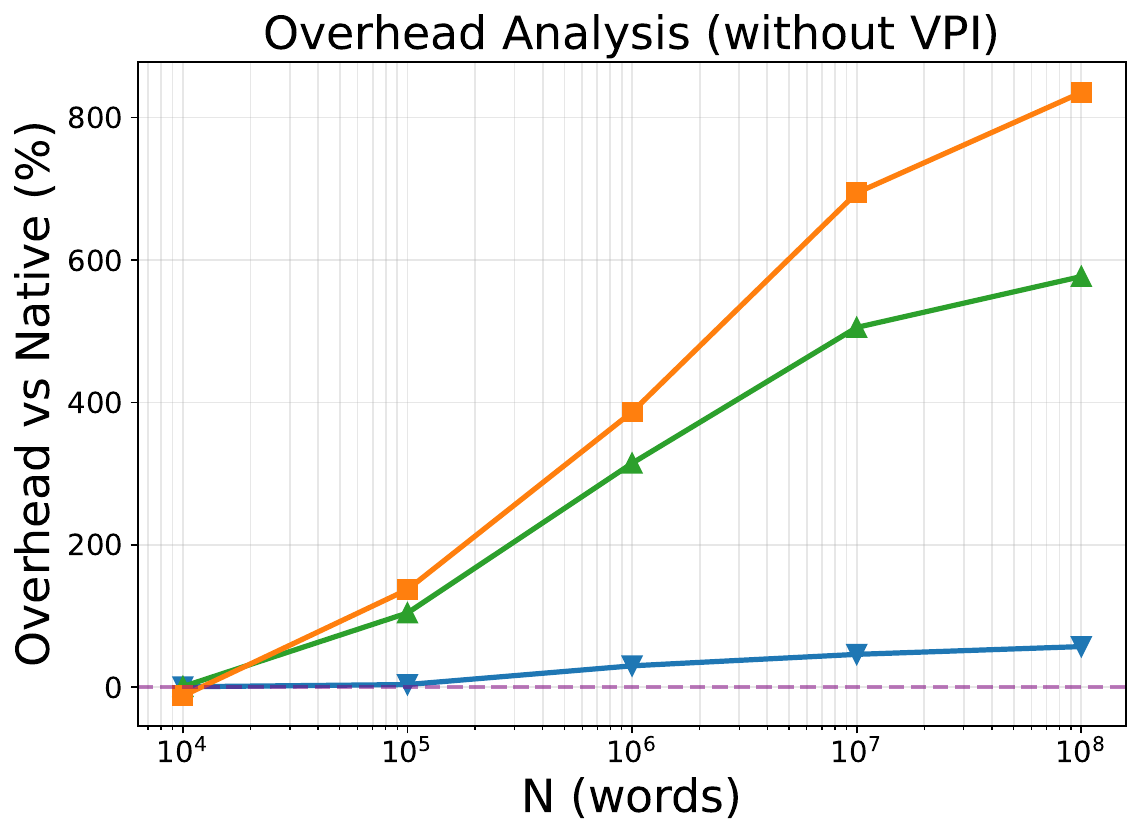} 
        \label{fig:overhead}
    \end{subfigure}
    \caption{Scalability Analysis: (Left) Total simulation time vs. Tensor size ($N$). (Right) Overhead percentage relative to Native C++ execution.}
    \label{fig:overhead_analysis}
    \vspace{-4pt}
\end{figure}

\noindent \textbf{Performance \& Trade-off Analysis.} 
Figure~\ref{fig:overhead_analysis} illustrates the scalability as the tensor size ($N$) increases from $10^4$ to $10^8$ words.
The \emph{vTen (Pure C++)} backend demonstrates near-native efficiency, tracking the Native C++ lower bound with $<$10\% overhead by completely eliminating the SystemVerilog boundary.
In contrast, the overhead of the \emph{vTen (SV-DPI)} backend increases linearly due to the inevitable context-switching cost (SV $\leftrightarrow$ C), yet it still outperforms the legacy \emph{File I/O} approach which suffers from prohibitive disk latency.
By fundamentally decoupling data transport (Shared Memory) from control (Kernel-Granular Batching), vTen effectively eliminates these IPC bottlenecks, establishing itself as a scalable architecture for verifying
data-intensive 3D workloads.

\subsection{End-to-End Performance Analysis}

\setlength{\textfloatsep}{2pt plus 1pt minus 1pt}
\setlength{\floatsep}{6pt plus 1pt minus 1pt}
\setlength{\intextsep}{6pt plus 1pt minus 1pt}
\setlength{\abovecaptionskip}{3pt}
\setlength{\belowcaptionskip}{-10pt}
\begin{figure}[t]
    \centering
    \includegraphics[width=1.0\linewidth]{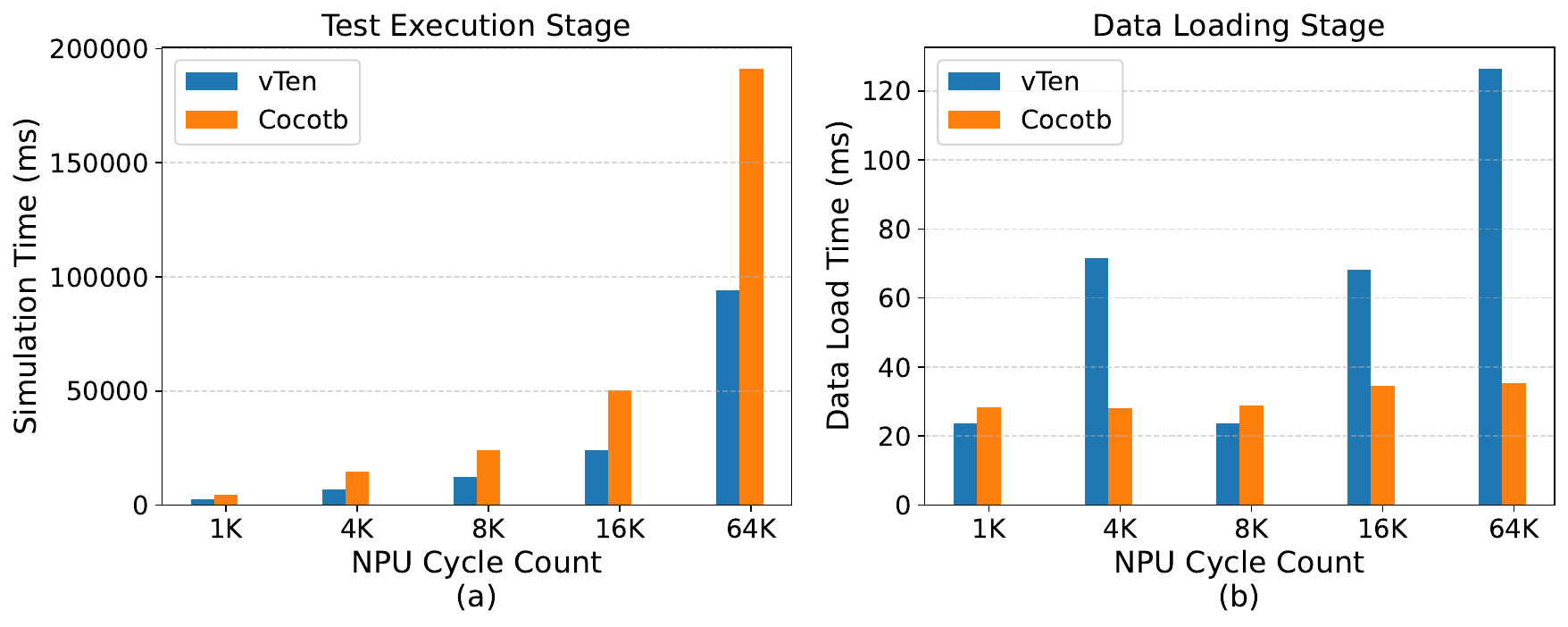} 
    \caption{Latency Comparison between vTen and Cocotb: vTen (blue) achieves a 2$\times$ performance increase over Cocotb (orange) in the Test Execution stage (a). vTen incurs a higher latency in the Data Loading stage (b) compared to Cocotb, though the overhead is minimal compared to (a). }
    \label{fig:performance_comparison}
    \vspace{-1pt}
\end{figure}

This experiment evaluates the end-to-end performance on the 3D U-Net case study defined in Section 4.1.
Figure~\ref{fig:performance_comparison}(a) shows that vTen achieves a 2$\times$ speedup in Test Execution ($T_{exec}$) compared to Cocotb across workloads ranging from 1K to 64K NPU cycles.
Since the DUT execution itself is identical, this gain is attributed entirely to the elimination of VPI-induced context switching required for clock advancement.
In Cocotb, the Python interpreter must wake up every cycle to drive the clock and signals, creating a dominant overhead that scales with simulation time ($O(C)$).
vTen removes this bottleneck by offloading clock generation and batched transactions to the backend. This architectural shift effectively neutralizes the cumulative latency incurred by massive repetitive operations (Challenge 1).

While Figure~\ref{fig:performance_comparison}(b) indicates higher loading latency ($T_{load}$) for vTen, this results from \textit{front-loading} data transfer to Shared Memory, whereas Cocotb defers signal driving to the execution phase.
This one-time setup effectively decouples transport from execution, enabling the performance gains demonstrated in Figure~\ref{fig:performance_comparison}(a).
Crucially, this upfront cost remains negligible (<1\% of total latency), ensuring that transport does not bottleneck the overall simulation.

\subsection{Verification Productivity Analysis}
Addressing Challenge 2 (System Integration) with heterogeneous IPs (e.g., DMA, MAC), vTen achieves a 92.2\% reduction in verification-specific code (28 vs. 360 lines). 
By encapsulating golden references within Kernel Objects, vTen facilitates hierarchical composition, allowing users to simply aggregate existing unit specifications instead of rewriting top-level logic. 
Consequently, this approach eliminates 242 lines of manual serialization and AXI4 sequencing required in Cocotb. 
Including the shared golden model, vTen demonstrates a 60.3\% overall reduction (261 vs. 658 lines), proving that Kernel Abstraction resolves integration bottlenecks via modular reuse.
\section{Related Work}
Existing hardware verification frameworks follow three approaches. 
Simulation acceleration includes hard\-ware-based solutions (Fire\-Sim\cite{firesim}, Diff\-Test-H\cite{difftest}, ASH\cite{ash}) that leverage FPGA/\allowbreak emu\-la\-tor platforms for faster 
simulation, and soft\-ware opti\-mi\-za\-tions (Veri\-lua\cite{verilua}) that improve simulator per\-for\-mance.
Design-verification languages (Chisel\cite{chisel}, PyMTL3\cite{pymtl3}) integrate verification capabilities directly into hardware construction languages, enabling unified design and test flows.
Portable verification frameworks (cocotb\cite{cocotb}, fault\cite{fault}) provide language-agnostic interfaces for test reuse across different simulators and formal tools. 
However, these approaches face fundamental limitations: hardware acceleration is primarily suited for SoC-level verification with significant setup overhead, design-specific languages create fragmented toolchains, and portable frameworks sacrifice performance for abstraction. 
These limitations are particularly acute for AI accelerator verification, which requires handling massive tensor datasets and complex data serialization---motivating vTen's compiled verification paradigm that eliminates per-transaction overhead while enhancing verification productivity.

\section{Conclusion}
In this paper, we propose vTen, a data-centric verification framework designed to bridge the semantic gap in Domain-Specific Accelerator (DSA) verification. 
By decoupling verification intent from execution mechanics through declarative DSL and kernel-granular batching, vTen effectively eliminates the synchronization bottlenecks found in traditional methodologies. 
Evaluation on a 3D U-Net accelerator demonstrates that vTen achieves a 2x performance improvement in test execution and a 60.3\% reduction in code complexity compared to Cocotb\cite{cocotb}.
These evaluation results highlight vTen's capability to provide a scalable, backend-agnostic verification solution, lowering entry barriers and accelerating agile development for next-generation AI hardware.

\begin{acks}
This research was supported in part by the ITRC (Information Technology Research Center) support program (IITP-2026-2020-0-01461) supervised by the IITP, funded by the MSIT (Ministry of Science and ICT), Korea, and in part by the Graduate School of Artificial Intelligence Semiconductor (IITP-2026-RS-2023-00256081) grant funded by the MSIT.
The EDA tool was supported by the IC Design Education Center (IDEC), Korea.
\end{acks}

\bibliographystyle{ACM-Reference-Format}
\bibliography{reference}


\begin{thebibliography}{24}


\ifx \showCODEN    \undefined \def \showCODEN     #1{\unskip}     \fi
\ifx \showISBNx    \undefined \def \showISBNx     #1{\unskip}     \fi
\ifx \showISBNxiii \undefined \def \showISBNxiii  #1{\unskip}     \fi
\ifx \showISSN     \undefined \def \showISSN      #1{\unskip}     \fi
\ifx \showLCCN     \undefined \def \showLCCN      #1{\unskip}     \fi
\ifx \shownote     \undefined \def \shownote      #1{#1}          \fi
\ifx \showarticletitle \undefined \def \showarticletitle #1{#1}   \fi
\ifx \showURL      \undefined \def \showURL       {\relax}        \fi
\providecommand\bibfield[2]{#2}
\providecommand\bibinfo[2]{#2}
\providecommand\natexlab[1]{#1}
\providecommand\showeprint[2][]{arXiv:#2}

\bibitem[uvm(2020)]%
        {uvm}
 \bibinfo{year}{2020}\natexlab{}.
\newblock \showarticletitle{{IEEE} Standard for Universal Verification Methodology Language Reference Manual}.
\newblock \bibinfo{journal}{\emph{IEEE Std 1800.2-2020 (Revision of IEEE Std 1800.2-2017)}} (\bibinfo{year}{2020}), \bibinfo{pages}{1--458}.
\newblock
\href{https://doi.org/10.1109/IEEESTD.2020.9195920}{doi:\nolinkurl{10.1109/IEEESTD.2020.9195920}}


\bibitem[Abadi et~al\mbox{.}(2016)]%
        {tensorflow}
\bibfield{author}{\bibinfo{person}{Mart{\'\i}n Abadi}, \bibinfo{person}{Paul Barham}, \bibinfo{person}{Jianmin Chen}, \bibinfo{person}{Zhifeng Chen}, \bibinfo{person}{Andy Davis}, \bibinfo{person}{Jeffrey Dean}, \bibinfo{person}{Matthieu Devin}, \bibinfo{person}{Sanjay Ghemawat}, \bibinfo{person}{Geoffrey Irving}, \bibinfo{person}{Michael Isard}, {et~al\mbox{.}}} \bibinfo{year}{2016}\natexlab{}.
\newblock \showarticletitle{{TensorFlow}: a system for Large-Scale machine learning}. In \bibinfo{booktitle}{\emph{12th USENIX symposium on operating systems design and implementation (OSDI 16)}}. \bibinfo{pages}{265--283}.
\newblock


\bibitem[Bachrach et~al\mbox{.}(2012)]%
        {chisel}
\bibfield{author}{\bibinfo{person}{Jonathan Bachrach}, \bibinfo{person}{Huy Vo}, \bibinfo{person}{Brian Richards}, \bibinfo{person}{Yunsup Lee}, \bibinfo{person}{Andrew Waterman}, \bibinfo{person}{Rimas Avi{\v{z}}ienis}, \bibinfo{person}{John Wawrzynek}, {and} \bibinfo{person}{Krste Asanovi{\'c}}.} \bibinfo{year}{2012}\natexlab{}.
\newblock \showarticletitle{Chisel: constructing hardware in a scala embedded language}. In \bibinfo{booktitle}{\emph{Proceedings of the 49th annual design automation conference}}. \bibinfo{pages}{1216--1225}.
\newblock


\bibitem[Cai et~al\mbox{.}(2025)]%
        {verilua}
\bibfield{author}{\bibinfo{person}{Ye Cai}, \bibinfo{person}{Chuyu Zheng}, \bibinfo{person}{Wei He}, {and} \bibinfo{person}{Dan Tang}.} \bibinfo{year}{2025}\natexlab{}.
\newblock \showarticletitle{Verilua: An Open Source Versatile Framework for Efficient Hardware Verification and Analysis Using LuaJIT}. In \bibinfo{booktitle}{\emph{2025 Design, Automation \& Test in Europe Conference (DATE)}}. IEEE, \bibinfo{pages}{1--2}.
\newblock


\bibitem[Elsabbagh et~al\mbox{.}(2023)]%
        {ash}
\bibfield{author}{\bibinfo{person}{Fares Elsabbagh}, \bibinfo{person}{Shabnam Sheikhha}, \bibinfo{person}{Victor~A Ying}, \bibinfo{person}{Quan~M Nguyen}, \bibinfo{person}{Joel~S Emer}, {and} \bibinfo{person}{Daniel Sanchez}.} \bibinfo{year}{2023}\natexlab{}.
\newblock \showarticletitle{Accelerating rtl simulation with hardware-software co-design}. In \bibinfo{booktitle}{\emph{Proceedings of the 56th Annual IEEE/ACM International Symposium on Microarchitecture}}. \bibinfo{pages}{153--166}.
\newblock


\bibitem[Foster(2024)]%
        {wilson2024}
\bibfield{author}{\bibinfo{person}{Harry~D. Foster}.} \bibinfo{year}{2024}\natexlab{}.
\newblock \bibinfo{title}{{Wilson} {Research} {Group} {IC}/{ASIC} functional verification trend report}.
\newblock
\urldef\tempurl%
\url{https://resources.sw.siemens.com/en-US/white-paper-2024-wilson-research-group-ic-asic-functional-verification-trend-report/}
\showURL{%
\tempurl}


\bibitem[Genc et~al\mbox{.}(2021)]%
        {gemmini}
\bibfield{author}{\bibinfo{person}{Hasan Genc}, \bibinfo{person}{Seah Kim}, \bibinfo{person}{Alon Amid}, \bibinfo{person}{Ameer Haj-Ali}, \bibinfo{person}{Vighnesh Iyer}, \bibinfo{person}{Pranav Prakash}, \bibinfo{person}{Jerry Zhao}, \bibinfo{person}{Daniel Grubb}, \bibinfo{person}{Harrison Liew}, \bibinfo{person}{Howard Mao}, {et~al\mbox{.}}} \bibinfo{year}{2021}\natexlab{}.
\newblock \showarticletitle{Gemmini: Enabling systematic deep-learning architecture evaluation via full-stack integration}. In \bibinfo{booktitle}{\emph{2021 58th ACM/IEEE Design Automation Conference (DAC)}}. IEEE, \bibinfo{pages}{769--774}.
\newblock


\bibitem[He et~al\mbox{.}(2015)]%
        {he_resnet}
\bibfield{author}{\bibinfo{person}{Kaiming He}, \bibinfo{person}{Xiangyu Zhang}, \bibinfo{person}{Shaoqing Ren}, {and} \bibinfo{person}{Jian Sun}.} \bibinfo{year}{2015}\natexlab{}.
\newblock \showarticletitle{Deep Residual Learning for Image Recognition}.
\newblock \bibinfo{journal}{\emph{CoRR}}  \bibinfo{volume}{abs/1512.03385} (\bibinfo{year}{2015}).
\newblock
\showeprint[arXiv]{1512.03385}
\urldef\tempurl%
\url{http://arxiv.org/abs/1512.03385}
\showURL{%
\tempurl}


\bibitem[Hong et~al\mbox{.}(2022)]%
        {dfx_hcs2022}
\bibfield{author}{\bibinfo{person}{Seongmin Hong}, \bibinfo{person}{Seungjae Moon}, \bibinfo{person}{Junsoo Kim}, \bibinfo{person}{Sungjae Lee}, \bibinfo{person}{Minsub Kim}, \bibinfo{person}{Dongsoo Lee}, {and} \bibinfo{person}{Joo-Young Kim}.} \bibinfo{year}{2022}\natexlab{}.
\newblock \showarticletitle{DFX: A Low-latency Multi-FPGA Appliance for Accelerating Transformer-based Text Generation}. In \bibinfo{booktitle}{\emph{2022 IEEE Hot Chips 34 Symposium (HCS)}}. \bibinfo{pages}{1--17}.
\newblock
\href{https://doi.org/10.1109/HCS55958.2022.9895626}{doi:\nolinkurl{10.1109/HCS55958.2022.9895626}}


\bibitem[Jiang et~al\mbox{.}(2020)]%
        {pymtl3}
\bibfield{author}{\bibinfo{person}{Shunning Jiang}, \bibinfo{person}{Peitian Pan}, \bibinfo{person}{Yanghui Ou}, {and} \bibinfo{person}{Christopher Batten}.} \bibinfo{year}{2020}\natexlab{}.
\newblock \showarticletitle{{PyMTL3}: A python framework for open-source hardware modeling, generation, simulation, and verification}.
\newblock \bibinfo{journal}{\emph{IEEE Micro}} \bibinfo{volume}{40}, \bibinfo{number}{4} (\bibinfo{year}{2020}), \bibinfo{pages}{58--66}.
\newblock


\bibitem[Jouppi et~al\mbox{.}(2023)]%
        {tpu}
\bibfield{author}{\bibinfo{person}{Norm Jouppi}, \bibinfo{person}{George Kurian}, \bibinfo{person}{Sheng Li}, \bibinfo{person}{Peter Ma}, \bibinfo{person}{Rahul Nagarajan}, \bibinfo{person}{Lifeng Nai}, \bibinfo{person}{Nishant Patil}, \bibinfo{person}{Suvinay Subramanian}, \bibinfo{person}{Andy Swing}, \bibinfo{person}{Brian Towles}, {et~al\mbox{.}}} \bibinfo{year}{2023}\natexlab{}.
\newblock \showarticletitle{Tpu v4: An optically reconfigurable supercomputer for machine learning with hardware support for embeddings}. In \bibinfo{booktitle}{\emph{Proceedings of the 50th annual international symposium on computer architecture}}. \bibinfo{pages}{1--14}.
\newblock


\bibitem[Karami et~al\mbox{.}(2025)]%
        {karami_ispass25_nongemm}
\bibfield{author}{\bibinfo{person}{Rachid Karami}, \bibinfo{person}{Sheng-Chun Kao}, {and} \bibinfo{person}{Hyoukjun Kwon}.} \bibinfo{year}{2025}\natexlab{}.
\newblock \bibinfo{title}{Understanding the Performance Horizon of the Latest ML Workloads with NonGEMM Workloads}.
\newblock
\showeprint[arxiv]{2404.11788}~[cs.AR]
\urldef\tempurl%
\url{https://arxiv.org/abs/2404.11788}
\showURL{%
\tempurl}


\bibitem[Karandikar et~al\mbox{.}(2018)]%
        {firesim}
\bibfield{author}{\bibinfo{person}{Sagar Karandikar}, \bibinfo{person}{Howard Mao}, \bibinfo{person}{Donggyu Kim}, \bibinfo{person}{David Biancolin}, \bibinfo{person}{Alon Amid}, \bibinfo{person}{Dayeol Lee}, \bibinfo{person}{Nathan Pemberton}, \bibinfo{person}{Emmanuel Amaro}, \bibinfo{person}{Colin Schmidt}, \bibinfo{person}{Aditya Chopra}, {et~al\mbox{.}}} \bibinfo{year}{2018}\natexlab{}.
\newblock \showarticletitle{{FireSim}: FPGA-accelerated cycle-exact scale-out system simulation in the public cloud}. In \bibinfo{booktitle}{\emph{2018 ACM/IEEE 45th Annual International Symposium on Computer Architecture (ISCA)}}. IEEE, \bibinfo{pages}{29--42}.
\newblock


\bibitem[Kim et~al\mbox{.}(2024)]%
        {kim_tcasi_2024}
\bibfield{author}{\bibinfo{person}{Minsik Kim}, \bibinfo{person}{Kyoungseok Oh}, \bibinfo{person}{Youngmock Cho}, \bibinfo{person}{Hojin Seo}, \bibinfo{person}{Xuan~Truong Nguyen}, {and} \bibinfo{person}{Hyuk-Jae Lee}.} \bibinfo{year}{2024}\natexlab{}.
\newblock \showarticletitle{A Low-Latency FPGA Accelerator for YOLOv3-Tiny With Flexible Layerwise Mapping and Dataflow}.
\newblock \bibinfo{journal}{\emph{IEEE Transactions on Circuits and Systems I: Regular Papers}} \bibinfo{volume}{71}, \bibinfo{number}{3} (\bibinfo{year}{2024}), \bibinfo{pages}{1158--1171}.
\newblock
\href{https://doi.org/10.1109/TCSI.2023.3335949}{doi:\nolinkurl{10.1109/TCSI.2023.3335949}}


\bibitem[Lee et~al\mbox{.}(2025)]%
        {lee_icce_2024}
\bibfield{author}{\bibinfo{person}{Keehyuk Lee}, \bibinfo{person}{Mincheol Cha}, \bibinfo{person}{Soosung Kim}, \bibinfo{person}{Xuan~Truong Nguyen}, {and} \bibinfo{person}{Hyuk-Jae Lee}.} \bibinfo{year}{2025}\natexlab{}.
\newblock \showarticletitle{A Winograd-Convolution-Based Accelerator on FPGA for Real-Time Object Detection with Effective On-Chip Buffer Access Patterns}. In \bibinfo{booktitle}{\emph{2025 IEEE International Conference on Consumer Electronics (ICCE)}}. \bibinfo{pages}{1--5}.
\newblock
\href{https://doi.org/10.1109/ICCE63647.2025.10930191}{doi:\nolinkurl{10.1109/ICCE63647.2025.10930191}}


\bibitem[Nguyen et~al\mbox{.}(2022)]%
        {nguyen_tcasi_2022}
\bibfield{author}{\bibinfo{person}{Duy~Thanh Nguyen}, \bibinfo{person}{Hyeonseung Je}, \bibinfo{person}{Tuan~Nghia Nguyen}, \bibinfo{person}{Soojung Ryu}, \bibinfo{person}{Kyujoong Lee}, {and} \bibinfo{person}{Hyuk-Jae Lee}.} \bibinfo{year}{2022}\natexlab{}.
\newblock \showarticletitle{ShortcutFusion: From Tensorflow to FPGA-Based Accelerator With a Reuse-Aware Memory Allocation for Shortcut Data}.
\newblock \bibinfo{journal}{\emph{IEEE Transactions on Circuits and Systems I: Regular Papers}} \bibinfo{volume}{69}, \bibinfo{number}{6} (\bibinfo{year}{2022}), \bibinfo{pages}{2477--2489}.
\newblock
\href{https://doi.org/10.1109/TCSI.2022.3153288}{doi:\nolinkurl{10.1109/TCSI.2022.3153288}}


\bibitem[Paszke et~al\mbox{.}(2019)]%
        {pytorch}
\bibfield{author}{\bibinfo{person}{Adam Paszke}, \bibinfo{person}{Sam Gross}, \bibinfo{person}{Francisco Massa}, \bibinfo{person}{Adam Lerer}, \bibinfo{person}{James Bradbury}, \bibinfo{person}{Gregory Chanan}, \bibinfo{person}{Trevor Killeen}, \bibinfo{person}{Zeming Lin}, \bibinfo{person}{Natalia Gimelshein}, \bibinfo{person}{Luca Antiga}, {et~al\mbox{.}}} \bibinfo{year}{2019}\natexlab{}.
\newblock \showarticletitle{Pytorch: An imperative style, high-performance deep learning library}.
\newblock \bibinfo{journal}{\emph{Advances in neural information processing systems}}  \bibinfo{volume}{32} (\bibinfo{year}{2019}).
\newblock


\bibitem[Redmon and Farhadi(2018)]%
        {yolov3}
\bibfield{author}{\bibinfo{person}{Joseph Redmon} {and} \bibinfo{person}{Ali Farhadi}.} \bibinfo{year}{2018}\natexlab{}.
\newblock \showarticletitle{YOLOv3: An Incremental Improvement}.
\newblock \bibinfo{journal}{\emph{CoRR}}  \bibinfo{volume}{abs/1804.02767} (\bibinfo{year}{2018}).
\newblock
\showeprint[arXiv]{1804.02767}
\urldef\tempurl%
\url{http://arxiv.org/abs/1804.02767}
\showURL{%
\tempurl}


\bibitem[Ronneberger et~al\mbox{.}(2015)]%
        {unet}
\bibfield{author}{\bibinfo{person}{Olaf Ronneberger}, \bibinfo{person}{Philipp Fischer}, {and} \bibinfo{person}{Thomas Brox}.} \bibinfo{year}{2015}\natexlab{}.
\newblock \showarticletitle{U-Net: Convolutional Networks for Biomedical Image Segmentation}.
\newblock \bibinfo{journal}{\emph{CoRR}}  \bibinfo{volume}{abs/1505.04597} (\bibinfo{year}{2015}).
\newblock
\showeprint[arXiv]{1505.04597}
\urldef\tempurl%
\url{http://arxiv.org/abs/1505.04597}
\showURL{%
\tempurl}


\bibitem[Rosser(2018)]%
        {cocotb}
\bibfield{author}{\bibinfo{person}{Benjamin~John Rosser}.} \bibinfo{year}{2018}\natexlab{}.
\newblock \showarticletitle{Cocotb: a Python-based digital logic verification framework}. In \bibinfo{booktitle}{\emph{Micro-electronics Section seminar. CERN, Geneva, Switzerland}}.
\newblock


\bibitem[Truong et~al\mbox{.}(2020)]%
        {fault}
\bibfield{author}{\bibinfo{person}{Lenny Truong}, \bibinfo{person}{Steven Herbst}, \bibinfo{person}{Rajsekhar Setaluri}, \bibinfo{person}{Makai Mann}, \bibinfo{person}{Ross Daly}, \bibinfo{person}{Keyi Zhang}, \bibinfo{person}{Caleb Donovick}, \bibinfo{person}{Daniel Stanley}, \bibinfo{person}{Mark Horowitz}, \bibinfo{person}{Clark Barrett}, {et~al\mbox{.}}} \bibinfo{year}{2020}\natexlab{}.
\newblock \showarticletitle{fault: A python embedded domain-specific language for metaprogramming portable hardware verification components}. In \bibinfo{booktitle}{\emph{International Conference on Computer Aided Verification}}. Springer, \bibinfo{pages}{403--414}.
\newblock


\bibitem[Wang et~al\mbox{.}(2021)]%
        {spatten}
\bibfield{author}{\bibinfo{person}{Hanrui Wang}, \bibinfo{person}{Zhekai Zhang}, {and} \bibinfo{person}{Song Han}.} \bibinfo{year}{2021}\natexlab{}.
\newblock \showarticletitle{Spatten: Efficient sparse attention architecture with cascade token and head pruning}. In \bibinfo{booktitle}{\emph{2021 IEEE International Symposium on High-Performance Computer Architecture (HPCA)}}. IEEE, \bibinfo{pages}{97--110}.
\newblock


\bibitem[You et~al\mbox{.}(2025)]%
        {difftest}
\bibfield{author}{\bibinfo{person}{Kunlin You}, \bibinfo{person}{Yinan Xu}, \bibinfo{person}{Kehan Feng}, \bibinfo{person}{Luoshan Cai}, \bibinfo{person}{Yaoyang Zhou}, {and} \bibinfo{person}{Yungang Bao}.} \bibinfo{year}{2025}\natexlab{}.
\newblock \showarticletitle{DiffTest-H: Toward Semantic-Aware Communication in Hardware-Accelerated Processor Verification}. In \bibinfo{booktitle}{\emph{Proceedings of the 58th IEEE/ACM International Symposium on Microarchitecture{\textregistered}}}. \bibinfo{pages}{1462--1476}.
\newblock


\bibitem[Özgün Çiçek et~al\mbox{.}(2016)]%
        {20163dunet}
\bibfield{author}{\bibinfo{person}{Özgün Çiçek}, \bibinfo{person}{Ahmed Abdulkadir}, \bibinfo{person}{Soeren~S. Lienkamp}, \bibinfo{person}{Thomas Brox}, {and} \bibinfo{person}{Olaf Ronneberger}.} \bibinfo{year}{2016}\natexlab{}.
\newblock \bibinfo{title}{3D U-Net: Learning Dense Volumetric Segmentation from Sparse Annotation}.
\newblock
\showeprint[arxiv]{1606.06650}~[cs.CV]
\urldef\tempurl%
\url{https://arxiv.org/abs/1606.06650}
\showURL{%
\tempurl}


\end{thebibliography}

\end{document}